\newcounter{myctr}

\documentclass{ws-ijqi}

\begin{document}

\markboth{H.-T. Elze et al.}
{A path integral for classical dynamics, entanglement, and Jaynes-Cummings model at ...}

\catchline{}{}{}{}{}

\title{A PATH INTEGRAL FOR CLASSICAL DYNAMICS, \\ ENTANGLEMENT, AND JAYNES-CUMMINGS MODEL \\  
AT THE QUANTUM-CLASSICAL DIVIDE  
}

\author{HANS-THOMAS ELZE, GIOVANNI GAMBAROTTA AND FABIO VALLONE 
}

\address{Dipartimento di Fisica ``Enrico Fermi'', Universit\`a di Pisa, \\ 
Largo Pontecorvo 3, I-56127 Pisa, Italia
\\ elze@df.unipi.it}



\maketitle

\begin{history}
\received{\today}
\end{history}

\begin{abstract}
The Liouville equation  
differs from the von~Neumann equation `only' by a characteristic superoperator.  
We demonstrate this for Hamiltonian dynamics, in general, and for 
the Jaynes-Cummings model, in particular. -- Employing superspace (instead of 
Hilbert space), we describe 
time evolution of density matrices in terms of path integrals which are formally 
identical for quantum and classical mechanics. They only differ by the interaction 
contributing to the action. 
This allows to import tools developed for Feynman path integrals, in order to 
deal with superoperators instead of quantum mechanical commutators 
in real time evolution. Perturbation 
theory is derived. Besides applications in classical statistical physics, 
the ``classical path integral'' and the parallel study of classical and quantum evolution 
indicate new aspects of (dynamically assisted) entanglement (generation).   
Our findings suggest to distinguish {\it intra-} from {\it inter-space entanglement}.  
\end{abstract}


\section{Introduction}	
The quantum-classical divide has been intensely studied in recent years 
with profound impact on various areas of 
research~\cite{KieferEtAl,Zurek,Schlosshauer}. In particular, this concerns  
the foundations of quantum mechanics, new quantum technologies (quantum information 
processing, precision measurements, designer materials, etc.), recent observations of 
quantum coherent processes in biology, and, 
last not least, unresolved issues surrounding ``quantum gravity''~\footnote{I.e., the 
conflict between quantum mechanics necessitating an external time and 
diffeomorphism invariance in general relativity, for example, which defies 
its existence. Despite its great successes in describing 
the statistical aspects of experiments, quantum theory itself presents  
problems of interpretation, which are brought to the forefront in    
quantum cosmology. They arise from its indeterministic features 
and are clearly seen, for example, in the measurement problem.}. 

Not surprisingly, these modern topics, which touch the foundations of 
quantum mechanics in one way or another, increase the impetus 
to try to reconstruct and to better understand the emergence of quantum 
mechanics from simpler dynamical structures beneath or more profound theoretical 
principles.   

Indeed, there is a growing number of deterministic models of quantum mechanical objects 
which are based on conjectured fundamental information loss, coarse graining,   
or dissipation 
mechanisms~\cite{tHooft06,Elze05,Blasone05,Adler,Smolin,Vitiello01,Isidro08,Wetterich08};  
see Refs.\,\cite{Elze09,tHooft09} for most recent arguments. -- 
We recall that 't Hooft's existence theorem~\cite{tHooft07} shows that  
the evolution of all quantum mechanical objects that are characterized by a finite 
dimensional Hilbert space can be captured by a dissipative process. This holds  
also for objects that are described by a set 
of mutually commuting Hermitean operators~\cite{Elze08}.  
  
However, a theory is lacking that would generally explain the emergence of quantum features of 
common objects, at the scales where they are observed. 

In order to make progress in these matters, it may be useful to further 
examine the quantum-classical divide. 
Presently, we look more carefully 
into the common as well as the distinctive features of classical and quantum 
dynamics, as described by the Liouville and the von~Neumann equations, respectively.   

We will derive a new path integral representation of the propagator 
for density matrices in the classical theory. It is identical with 
the usual one at the kinematic level, employing the Feynman propagator of quantum mechanics~\cite{Schulman}; 
this allows `external sources' in the relevant action that are 
coupled to terms linear or quadratic in the generalized coordinates. 
Yet the interaction part differs in a characteristic way~\footnote{Another path 
integral for classical mechanics exists, which  
implements Hamilton's equations as constraints, see Refs.~\cite{Blasone05} and 
references there. In this approach, an analogy with quantum mechanics consists 
in the path integral as such, yet its integrand bears no resemblance.}. 
The new formalism based on superoperators will be presented and illustrated here by 
perturbation theory applied to an anharmonic oscillator. 

Similarly, as a case study, we will re-derive the Jaynes-Cummings 
model~\cite{JaynesCummings} -- 
the well-known benchmark model of quantum optics and cavity QED -- 
based on classical 
dynamics described by a Liouville equation. Thus, when applied to the two-level 
dynamics of Rydberg atoms coupled to one mode of the photon field in a suitably tuned cavity, 
we find surprisingly that it is ``almost classical'', with quantum 
and classical dynamics differing by a characteristic superoperator.   

We conclude by pointing out some interesting problems, concerning the preparation of 
entangled states, in particular. Here, the 
quantum-classical divide shows new aspects, which may help to further unravel the hidden  
dynamics that must be involved when it is crossed -- be it in the ``classical limit'' 
or following axiomatic ``quantization rules''.    
  
\section{Hamiltonian dynamics and the Liouville superoperator}

To begin with, we will consider an object  
with a single continuous degree of freedom. We will treat an atom interacting 
with the electromagnetic field in a later chapter, 
while a relativistic field theory has been studied elsewhere~\cite{Elze07}. 
  
Let us assume that there are only conservative forces and that Hamilton's equations 
are determined by the generic Hamiltonian function:  
\begin{equation}\label{HamiltonianF} 
H(x,p):=\frac{1}{2}p^2+V(x) 
\;\;, \end{equation} 
defined in terms of generalized coordinate $x$ and momentum $p$ (a mass parameter will 
be inserted in due time, but is omitted here for simplicity), and where  
$V(x)$ denotes the potential. -- 
An ensemble of such objects, for example, 
following trajectories with different initial conditions, is described by 
a distribution function $\rho$ in phase space, i.e., by the probability 
$\rho(x,p;t)\mbox{d}x\mbox{d}p$ to find a member of the ensemble in an infinitesimal 
volume at point $(x,p)$. This distribution evolves according to the {\it Liouville equation}: 
\begin{equation}\label{LiouvilleEq} 
-\partial_t\rho =\frac{\partial H}{\partial p}\cdot\frac{\partial \rho}{\partial_x}
-\frac{\partial H}{\partial x}\cdot\frac{\partial \rho}{\partial_p}
=\big\{p\partial_x-V'(x)\partial_p\big\}\rho  
\;\;, \end{equation} 
with $V'(x):=\mbox{d}V(x)/\mbox{d}x$. -- We recall that the relative minus sign 
in the Poisson bracket, or between terms here,   
reflects the symplectic phase space symmetry. This will translate into the familiar 
commutator structure in Eq.\,(\ref{Schroed}).   

A Fourier transformation, $\rho (x,p;t)=\int\mbox{d}y\;\mbox{e}^{-ipy}\rho (x,y;t)$, 
replaces the Liouville equation by: 
\begin{equation}\label{LFourier}  
i\partial_t\rho =\big\{-\partial_y\partial_x+yV'(x)\big\}\rho  
\;\;, \end{equation} 
without changing the symbol for the distribution function,  
whenever changing variables.  
Thus, momentum is eliminated in favour of {\it doubling} the number of coordinates. Finally, 
with the transformation: 
\begin{equation}\label{coordtrans} 
Q:=x+y/2\;\;,\;\;\;q:=x-y/2  
\;\;, \end{equation} 
we obtain the Liouville equation in the form: 
\begin{eqnarray}\label{Schroed} 
i\partial_t\rho &=&\big\{ \hat H_Q-\hat H_q+{\cal E}(Q,q)\big\}\rho 
\;\;, \\ [1ex] \label{HX} 
\hat H_\chi &:=&-\frac{1}{2}\partial_\chi ^{\;2}+V(\chi )\;\;, 
\;\;\;\mbox{for}\;\;\chi =Q,q 
\;\;, \\ [1ex] \label{I} 
{\cal E}(Q,q)&:=&(Q-q)V'(\frac{Q+q}{2})
-V(Q)+V(q)\;=\;-{\cal E}(q,Q)
\;\;. \end{eqnarray}  
We remark that the presented reformulation of classical dynamics 
is rather independent of the number of degrees of freedom. 
It applies to matrix valued as well as to   
Grassmann valued variables, representing the ``pseudoclassical'' fermion  
fields introduced by Casalbuoni and by Berezin and Marinov. 
Field theories require a classical 
functional formalism~\cite{Elze05,Elze07}. 

Furthermore, the Eq.\,(\ref{Schroed}) appears as the {\it von\,Neumann equation} 
for a density operator $\hat \rho (t)$, 
considering $\rho (Q,q;t)$ as its matrix elements. 
We automatically recover the Hamiltonian operator $\hat H$  
related to the Hamiltonian function, Eq.\,(\ref{HamiltonianF}),  
as in quantum theory. 
However, an essential difference consists   
in the interaction ${\cal E}$ between 
{\it bra-} and {\it ket-} states. 
The Hilbert space and its dual here are coupled by 
a {\it superoperator}~\footnote{This superoperator is of a very    
specific form, which leads to the antisymmetry in 
Eq.\,(\ref{I}). It differs from a Lindblad superoperator, often  
obtained as a symmetric double commutator structure, 
in the case of open quantum mechanical 
systems~\cite{Diosi}.}. 

Since the interaction ${\cal E}$ is antisymmetric under 
$Q\leftrightarrow q$, the complete   
(Liouville) operator on the right-hand side of Eq.\,(\ref{Schroed}) has a symmetric spectrum 
with respect to zero and, in general, will not be bounded below. 
Therefore, with this coupling of the Hilbert space and its dual by the superoperator, 
corresponding to the absence of a stable ground state, 
our reformulation of Hamiltonian dynamics does not qualify as a quantum theory. 
Related observations were discussed, for example, in  
Refs.~\cite{tHooft06,Elze05,Blasone05,Vitiello01}. 

However, the following fact 
has been discussed in Refs.\,\cite{Elze09}: 
\begin{equation}\label{Ezero}
{\cal E}\equiv 0\;\;\Longleftrightarrow\;\;
\mbox{potential}\;V(x)\;\mbox{is constant, linear, or harmonic} 
\;, \end{equation} 
with an eye on the possibility of having   
quantum phenomena emerge due to discrete spacetime structure.  
Analogously, the vanishing of ${\cal E}$ in a field theory is equivalent   
with having massive or massless free fields, with or without external sources, 
and with or without bilinear couplings. Generally, in these cases, anharmonic 
forces or interactions are absent.    

In the following main parts of this work, we will study in more detail 
the classical Hamiltonian dynamics described by Eq.\,(\ref{Schroed}), or by its 
appropriate 
generalizations, and pay particular attention to the presence of the 
superoperator ${\cal E}$, when comparing with the von\,Neumann equation.  

Concluding this introductory section, we recall relevant aspects of  
the interpretation of the density operator $\hat\rho (t)$, 
which we invoked here. 

\subsection{Expectations, operators and the Born rule}

We begin with the normalization of the classical probability distribution: 
\begin{equation}\label{clnorm} 
1\stackrel{!}{=}\int\frac{\mbox{d}x\mbox{d}p}{2\pi}\;\rho (x,p;t)=\int\mbox{d}Q\mbox{d}q\;
\delta (Q-q)\rho (Q,q;t)=:\mbox{Tr}\;\hat\rho (t) 
\;\;, \end{equation}  
incorporating the transformations of Section\,2. Consider a    
complete set of orthonormal eigenfunctions of the operator $\hat H_\chi $ of Eq.\,(\ref{HX}), 
defined by $g_j(\chi ;t):=\mbox{exp}(-iE_jt)g_j(\chi )$ and 
$\hat H_\chi g_j(\chi )=E_jg_j(\chi )$, respectively, 
with a discrete spectrum, for simplicity. 
Then, we may expand $\rho$: 
\begin{equation}\label{fexpans} 
\rho (Q,q;t)=\sum_{j,k}\rho_{jk}(t)g_j(Q;t)g_k^*(q;t)
\;\;. \end{equation} 
Employing this, the normalization condition (\ref{clnorm}) can be stated as: 
\begin{equation}\label{clnorm1} 
1\stackrel{!}{=}\sum_{j,k}\rho_{jk}(t)\mbox{e}^{-i(E_j-E_k)t}\int\mbox{d}Q\;g_j(Q)g_k(Q)
=\sum_{j}\rho_{jj}(t)
\;\;. \end{equation} 
Since the classical phase space 
distribution is real, the expansion coefficients form 
a Hermitean matrix, $\rho_{ij}=\rho^\ast_{ji}$, which we also denote by $\hat\rho$.  
  
The {\it classical} expectation values are calculated as follows: 
\begin{eqnarray}\label{xexpect} 
\langle x\rangle :=\int\frac{\mbox{d}x\mbox{d}p}{2\pi}\;x\rho (x,p;t)
&=&\int\mbox{d}Q\mbox{d}q\;\delta (Q-q)\frac{Q+q}{2}\rho (Q,q;t)
\;\;, \\ [1ex] \label{xexpect1} 
&=:&\mbox{Tr}\;\big (\hat X\hat\rho (t)\big ) 
\;\;, \\ [1ex] \label{pexpect} 
\langle p\rangle :=\int\frac{\mbox{d}x\mbox{d}p}{2\pi}\;p\rho (x,p;t)
&=&\int\mbox{d}Q\mbox{d}q\;\delta (Q-q)(-i)\frac{\partial_Q-\partial_q}{2}\rho (Q,q;t)
\;\;, \\ [1ex] \label{pexpect1} 
&=:&\mbox{Tr}\;\big (\hat P\hat\rho (t)\big ) 
\;\;, \end{eqnarray}  
introducing the operators $\hat X$ and $\hat P$, with matrix elements 
$X(q,Q)=\delta (Q-q)(Q+q)/2$ and 
$P(q,Q)=-i\big (\delta (Q-q)\stackrel{\rightharpoondown}{\partial}_Q-
\stackrel{\leftharpoondown}{\partial}_q\delta (Q-q)\big )$ (derivatives act left 
or right, as indicated). 
Eliminating one of the two integrations in the above equations with the help 
of the $\delta$-functions and suitable partial integrations, these operators are 
recognized as the coordinate and momentum operators of quantum theory. 
 
Similarly, we find: 
\begin{equation}\label{noncommops} 
\int\frac{\mbox{d}x\mbox{d}p}{2\pi}\;xp\rho (x,p;t) = 
\frac{1}{2}\mbox{Tr}\;\big ((\hat X\hat P+\hat P\hat X)\hat\rho (t)\big ) 
\;\;, \end{equation} 
which constitutes an example of the 
symmetric Weyl ordering, when replacing classical phase space quantities by 
quantum operators. -- However, we remark that Hilbert space 
operators appear here by rewriting classical  
statistical formulae and {\it not} by following a quantization rule.   

The Eqs.\,(\ref{clnorm}), (\ref{xexpect})--(\ref{noncommops}) are in accordance 
with the interpretation of $\rho (Q,q;t)$ as matrix elements of a density operator $\hat\rho (t)$. -- 
However, there is an important {\it caveat}: The eigenvalues of normalized 
quantum mechanical density operators are usually constrained to lie between zero and one, 
corresponding to the interpretation as standard probabilities.  
This is not necessarily the case with the 
operator $\hat\rho $ obtained from a classical probability distribution. Similarly, 
the Wigner distribution -- obtained from the matrix elements of 
a quantum mechanical density operator by applying the   
transformations leading from $\rho (x,p)$ to $\rho(Q,q)$ in reverse -- generally, is not 
positive semi-definite on phase space, even though its marginal distributions are. 
Therefore, it does not qualify as a classical probability density.  

As we have indicated before, there is clearly a dynamical feature missing, which 
governs the crossing of the quantum-classical divide, if not done `by hand', as 
in any of the usual ``quantization prescriptions''. Last not least, this must 
establish the {\it Born rule} by eliminating negative probabilities or by 
leading to their satisfactory interpretation. 

\section{From Hilbert space to superspace}  

In this section, we reformulate the notions relevant for the dynamics of density 
operators, at which we arrived in the previous section, in a more convenient way, 
introducing the concept of {\it superspace}~\footnote{The notion of 
{\it superspace} here, at first sight, has little in common and should not be 
confused with the corresponding term relating to supersymmetry.}, also called   
{\it Liouville space} -- see Ref.\,\cite{superspace} for a concise presentation 
and numerous applications.    

Considering a physical object characterized by the Hamiltonian $\hat{H}$, as in quantum 
theory, we introduce a complete set of basis states, $\{ |j\rangle\}$ ($j=1,\dots ,N$), assuming 
that the relevant Hilbert space is $N$-dimensional. Then, taking matrix elements of the von\,Neumann equation, 
for example, we have:  
\begin{equation}\label{vN}
i \partial_t \rho_{jk} = [(\hat{H}\hat{\rho})_{jk}- (\hat{\rho}\hat{H})_{jk}]\;\;, \;\;\;  
j,k = 1,2,\dots, N
\;\;, \end{equation}
with a density matrix $\rho$ of $N^2$ elements. Which may be written as:   
\begin{equation}
i \partial_t \rho_{jk} = \sum_{l,m} {\cal L}_{jk,lm} \rho_{lm}
\;\;, \end{equation}
in terms of the {\it Liouville superoperator} $\hat {\cal L}$ defined by: 
\begin{equation}\label{LvN}
\hat {\cal L}_{jk,lm}:= H_{jl} \delta_{km} - H_{km}^\ast \delta_{jl} 
\;\;. \end{equation}
This definition suggests to introduce a space where the density operator is a vector, 
which is the Liouville space (or superspace). The dynamics of the density operator 
can then be more conveniently described, completely in parallel for classical 
and quantum mechanics, as we shall see. 

Given the Hilbert space, as above, the density operator can be expanded as:  
\begin{equation}
\hat{\rho}= \sum_{j,k} \rho_{jk} |j \rangle \langle k| 
\;\;. \end{equation}
We may think of the family of $N^2$ operators $|j\rangle \langle k|$, with $j,k=1,\cdots,N$, as a 
complete set of matrices, or vectors, and express the density operator as: 
\begin{equation}
|\rho \rangle= \sum_{j,k} \rho_{jk} |jk \rangle \rangle 
\;\;, \end{equation}
where the ``ket'' $|jk \rangle \rangle$ denotes the Liouville space {\it vector} representing the 
Hilbert space {\it operator} $|j\rangle \langle k|$. Similarly, we introduce a ``bra'' vector
$ \langle \langle jk|$ as the Hermitean conjugate of $|jk \rangle \rangle$.

In Liouville space, any operator $\hat{A}$ is represented by a vector and denoted by 
$|A \rangle \rangle $. It can be expanded as:  
\begin{equation} \label{opex}
| A \rangle \rangle= \sum_{j,k} A_{jk} |jk \rangle \rangle
\;\;, \end{equation}
where $A_{jk}$ are the usual matrix elements $\langle j| \hat{A}| k \rangle$. -- 
Furthermore, we can define a ``bra'' vector $ \langle \langle B |$, representing $\hat{B}^{\dag}$, 
and the scalar product of two operators: 
\begin{equation}
\langle \langle B| A \rangle \rangle :=\mbox{Tr} (\hat{B}^{\dag}\hat{A})   
\;\;. \end{equation}
Then, the following orthonormality condition holds: 
\begin{equation}\label{orthogonality} 
\langle \langle jk|mn \rangle \rangle =\mbox{Tr}(|k\rangle \langle j| m \rangle \langle n|)=
\delta_{kn} \delta_{jm} 
\;\;, \end{equation} 
which is analogous to $\langle j|k\rangle = \delta_{jk}$. -- Finally, consider the scalar product: 
\begin{equation}
\langle \langle jk| A \rangle \rangle =\mbox{Tr}(|k \rangle \langle j| \hat{A})=
\sum_{l} \langle l|k \rangle \langle j |\hat{A}| l \rangle= \langle j| \hat{A}|k \rangle \equiv A_{jk}
\;\;. \end{equation} 
Upon substitution in Eq.\,(\ref{opex}), this yields: 
\begin{equation}
|A \rangle \rangle = \sum_{j,k} |jk \rangle \rangle \langle \langle jk | A \rangle \rangle 
\;\;. \end{equation}
This is consistent with the following completeness relation in Liouville space: 
\begin{equation}\label{completeness}
\sum_{j,k} |jk \rangle \rangle \langle \langle jk | = {\mathbf 1} 
\;\;. \end{equation}
Following these considerations, it can be verified that Liouville
space is a linear space, in which the density operator $\hat{\rho}$ is a vector. 
In this space, a linear operator can be defined by: 
\begin{equation}
\hat{\cal F}:= \sum_{j,k,m,n} |jk \rangle \rangle \langle \langle jk |
\hat{\cal F}|mn \rangle \rangle \langle \langle mn| \equiv   
\sum_{j,k,m,n}{\cal F}_{jk,mn}|jk \rangle \rangle \langle \langle mn| 
\;\;, \end{equation} 
i.e., in terms of appropriate matrix elements. 

The importance of Liouville space for classical {\it and} quantum dynamics is that 
the Liouville and von\,Neumann equations, both, can be written in the form:     
\begin{equation}
i \partial_t \hat{\rho} = \hat {\cal L} \hat{\rho}
\;\;, \end{equation}
with an appropriate superoperator $\hat {\cal L}$, cf. 
Eqs.\,(\ref{Schroed})--(\ref{I}) and (\ref{vN})--({\ref{LvN}), respectively. 
Thus, there is a formal analogy (even isomorphism) between the structure of these equations 
and the Schr\"odinger equation, 
$i\partial_t \Psi = \hat{H} \Psi \;$. 

These observations suggest that techniques or formal results concerning the solution of 
the Schr\"odinger equation can be transferred to the case of the Liouville or 
von\,Neumann equations with the help of Liouville space notions. This concerns 
perturbation theory (and nonperturbative methods) as much as a path integral formulation, 
which we shall discuss in turn. 

First of all, we introduce the Liouville space evolution operator $\hat{\cal U}$ 
satisfying:  
\begin{equation}
i \partial_t \hat{\cal U}(t,t_0) = \hat{\cal L}(t)\hat{\cal U}(t,t_0) 
\;\;, \end{equation}
with the initial condition $\hat{\cal U}(t,t_0)=\mathbf{1}$. This implies:   
\begin{equation} \label{Levol}
\hat{\rho}(t) = \hat{\cal U}(t,t_0) \hat{\rho}(t_0)
\;\;, \end{equation} 
as the solution of the density operator equation of motion. 
For a time independent Liouville superoperator this yields: 
\newcommand{\Liu}{\hat{\cal L}}
\newcommand{\U}{\hat{\cal U}}
\newcommand{\Liui}{\hat{\cal L}_I}
\newcommand{\Ui}{\hat{\cal U}_I}
\begin{equation}
\U(t,t_0)= \exp \big (-i \Liu (t-t_0)\big )  
\;\;. \end{equation}
Thus, time evolution of the density matrix is implemented by a superoperator in Liouville 
space, while in Hilbert space the evolution is described by: 
\newcommand{\ro}{\hat{\rho}}
\begin{equation}\label{hevol}
\ro(t)= \hat{U}(t,t_0) \ro(t_0) \hat{U}^{\dag}(t,t_0) 
\;\;, \end{equation} 
with $\hat{U}(t,t_0):=\exp (-i\hat Ht)$. 
The Eqs.\,(\ref{Levol}) and (\ref{hevol}) represent the evolution of the same object, 
although in different spaces. -- For a time dependent Hamiltonian, we have instead: 
\begin{eqnarray}
&\;&\U(t,t_0)= \mbox{T}\exp\big (-i \int_{t_0}^t\mbox{d}\tau\; \Liu (\tau) \big )  
\\ [1ex] 
&\;&\;\;\; := 
1 + \sum_{n=1}^{\infty} (-i)^n \int_{t_0}^t\mbox{d}\tau_n \int_{t_0}^{\tau_n}\mbox{d}\tau_{n-1}
\;\dots \int_{t_0}^{\tau_2}\mbox{d}\tau_1\;\Liu (\tau_n)\Liu(\tau_{n-1})\dots \Liu (\tau_1)
\;, \end{eqnarray}
in terms of the time-ordered exponential.

For later purposes, we also introduce the ``interaction picture'' in Liouville space. 
Considering a Liouville operator which consists of two parts: 
\begin{equation}
\Liu\equiv \Liu_0(t)+ \Liu'(t)
\;\;, \end{equation}
we obtain the evolution operator in the following form: 
\begin{equation}
\U(t,t_0) = \U_0(t,t_0) \U_I(t,t_0)
\;\;, \end{equation}
with: 
\begin{equation}
\U_0(t,t_0)=\mbox{T} \exp \big (-i \int_{t_0}^t\mbox{d}\tau\;\Liu_0 (\tau) \big ) 
\;\;, \end{equation}
and:
\begin{equation}\label{UI}
\U_I(t,t_0)=\mbox{T} \exp \big (-i \int_{t_0}^t\mbox{d}\tau\;\Liu'_I(\tau) \big ) 
\;\;, \end{equation}
with $ \Liu'_I (\tau) := \U^{\dag}_0(\tau,t_0) \Liu'(\tau) \U_0 (\tau,t_0)$.
For an operator $\U_0$ that can be treated exactly, study of time evolution 
essentially concerns the operator $\U_I$ -- for this, Eq.\,(\ref{UI}) 
presents the starting point of perturbation theory (expanding the exponential). 

\section{The quantum/classical path integral for the propagator of density matrices}   

The technical ingredients needed for the Feynman path integral approach, 
for the derivation of quantum mechanical propagators in particular, are very 
well known. We import these here, especially from Ref.\,\cite{Schulman}, in order 
to derive a path integral for the propagator of density matrices based on 
the Liouville space formulation of the preceding Section\,3. 

Our derivation relies on the close formal similarity between the classical 
Liouville equation and the von\,Neumann equation on one hand side and the 
Schr\"odinger equation on the other, in an appropriate representation, as 
we have discussed~\footnote{In this chapter, we reinstate $\hbar$ explicitly.}.  

\subsection{Essentials of the Feynman path integral}

We recall that the (forward propagating) operator Green's function $\hat G$, 
\begin{equation}
\hat G(t,t_0)\equiv \theta (t-t_0)\exp \big (-i\hat H(t-t_0)/\hbar\big ) 
\;\;, \end{equation}  
allows one to write the solution of the time dependent Schr\"odinger equation as: 
$|\psi (t)\rangle =\hat G(t,t_0)|\psi (t_0)\rangle$. -- Correspondingly, 
for $t>t_0$, we have the coordinate space matrix elements: 
\begin{equation}
G(x,t;y,t_{0})=\langle x|\mbox{e}^{-i\hat H(t-t_0)/\hbar}|y\rangle
\;\;, \end{equation}  
from which one derives the path integral representation of these amplitudes, for a generic {\it Hamiltonian}  
$\hat H=\hat p^2/2m+V(\hat x)$, describing a particle of mass $m$ in an external potential $V$,   
cf. Eq.(\ref{HamiltonianF}), through the following steps~\cite{Schulman}: 
\begin{itemize}
\item Cut the time interval from $t_0$ to $t$ into a large number $N$ of equal pieces.  
\item Write the exponential of the Hamiltonian operator$\;\times\;$time as a product of identical factors, 
each factor representing the propagator for a small time interval $\propto 1/N$. 
\item Separate the kinetic and potential terms contributing to $\hat H$ in each one of these 
factors with the crucial help of the {\it Trotter product formula}. 
\item Alternatingly, insert complete sets of momentum and coordinate eigenstates, 
such as $\int\mbox{d}x\;|x\rangle\langle x |={\mathbf 1}$ (and correspondingly for 
momentum eigenstates) between 
the factors of exponentials involving either momentum or coordinate operators and 
evaluate the resulting Gaussian integrals over momentum variables. 
\item Realize that the obtained phases in the product of exponentials 
can be summed up to represent a discretized version of the {\it classical action} pertaining 
to the Hamiltonian function (corresponding to $\hat H$). 
\end{itemize}
Taking the limit $N\rightarrow\infty$ in the end, one obtains the following Feynman path integral 
representation of the amplitudes in question: 
\begin{eqnarray} \label{piprop} 
&\;&G(x,t;y,t_{0})\;= 
\\ [1ex] 
&\;&\lim_{N\rightarrow\infty}(\frac{m}{2\pi i\hbar\epsilon})^{N/2} 
\int\mbox{d}x_1\dots\mbox{d}x_{N-1}\;
\exp \Big (\frac{i\epsilon}{\hbar}
\sum_{j=0}^{N-1}\big (\frac{m}{2}(\frac{x_{j+1} -x_{j}}{\epsilon})^{2} -V(x_{j})
\big )\Big ) 
\nonumber \\ [1ex] \label{pathintegral1} 
&=:&\int{\cal D}x\;\exp\Big (\frac{i}{\hbar}\int_{t_0}^t\mbox{d}\tau\; 
\big (\frac{m}{2}\dot x^2 -V(x)\big )\Big) 
\\ [1ex] \label{pathintegral2} 
&\equiv&\int{\cal D}x\;\exp\big (\frac{i}{\hbar}S[\dot x,x]\big )
\;\;, \end{eqnarray} 
where $\epsilon :=(t-t_0)/N$, $\dot x:=\mbox{d}x/\mbox{d}\tau$ and 
where $S$ denotes the 
relevant classical action, which is to be evaluated for each one 
of the paths contributing to the integral, with the 
boundary conditions $x(t)=x$ and $x(t_0)=y$. 

\subsection{The Liouville space propagator as a path integral}
We are now ready to appreciate the economy of the Liouville space representation introduced 
in Section\,3.  
In particular, the formal solution of the classical Liouville equation 
as well as of the quantum mechanical von\,Neumann equation, both, can be written in the form: 
\begin{equation}
|\rho(t)\rangle\rangle =\mbox{e}^{-i{\cal \hat H}(t-t_0)/\hbar}|\rho(t_0)\rangle\rangle 
\;\;, \end{equation} 
where ${\cal \hat H}$ is the appropriate super-Hamiltonian. Generally, we have: 
\begin{equation}
\langle\langle Q,q|{\cal \hat H}|Q',q'\rangle\rangle =
\delta (Q-Q')\delta (q-q')\big (\hat H(Q)-\hat H(q)+{\cal E}(Q,q))
\;\;, \end{equation} 
where $\hat H$ denotes the appropriate Hamilton operator in coordinate representation, as 
indicated, which alone is relevant for the von\,Neumann equation, while ${\cal E}$ 
represents the additional superoperator for classical dynamics, cf. Section\,2. 

In order to solve the problem of time evolution in the present case, we need to know 
the (super)matrix elements entering the propagation equation: 
\begin{equation}\label{rhoprop}
\langle\langle Q,q|\rho(t)\rangle\rangle =\int\mbox{d}Q'\mbox{d}q'\;
\langle\langle Q,q|\mbox{e}^{-i{\cal \hat H}(t-t_0)/\hbar}|Q',q'\rangle\rangle\langle\langle Q',q'|\rho(t_0)\rangle\rangle
\;\;, \end{equation} 
which appears formally analogous to evolution of a state vector according the Schr\"odinger equation. 
Thus, not surprisingly, we go through the steps indicated in the preceding Section\,4.1, in order 
to construct the path integral representation of the propagator here. 

However, in this derivation, we have to pay attention to the crucial role of the 
Trotter product formula. It turns out that it can be generalized for our purposes,   
where superoperators present the new feature, in a straightforward way; the relevant 
definitions and details of the proof will be given 
elsewhere~\cite{FabThesis}.  

Rewriting the Eq.\,(\ref{rhoprop}) as: 
\begin{equation}\label{rhoprop1} 
\rho (Q,q;t) = \int\mbox{d}Q'\mbox{d}q'\;{\cal G}(Q,q;t|Q',q';t_0)\rho (Q',q';t_0) 
\;\;, \end{equation} 
our interest is to know the superpropagator ${\cal G}$. Next, we will follow the   
recipe to arrive at a path integral representation, as summarized above,  
in Section\,4.1. 
 
In particular, here we make use 
of suitably inserted complete sets of superspace vectors, such as: 
\begin{equation}\label{supercomplete} 
\int\mbox{d}Q\mbox{d}q\; |Q,q\rangle\rangle\langle\langle Q,q|={\mathbf 1} 
\;\;, \end{equation} 
and, correspondingly, for momentum space, cf. Eq.\,(\ref{completeness}).  
Using the plane wave relation between coordinate and momentum eigenfunctions, we also employ: 
\begin{equation}\label{planewaves}
\langle\langle P,p|=\frac{1}{2\pi\hbar }\int\mbox{d}Q\mbox{d}q\;
\exp\big (-\frac{i}{\hbar}(PQ-pq)\big )\langle\langle Q,q| 
\;\;. \end{equation} 
Furthermore, the orthogonality relation 
$\langle\langle Q,q|Q',q'\rangle\rangle =\delta (Q-Q')\delta(q-q')$, cf. 
Eq.\,(\ref{orthogonality}), implies: 
\begin{equation}\label{transf} 
\langle\langle P,p|Q,q\rangle\rangle =\frac{1}{2\pi\hbar }
\exp\big (-\frac{i}{\hbar}(PQ-pq)\big )
\;\;. \end{equation} 
Then, with all following steps of the derivation in parallel with the usual ones in  
quantum mechanics, it is straightforward 
to obtain in the present case~\cite{FabThesis}: 
\begin{equation}\label{superpropagator} 
{\cal G}(Q_f,q_f;t|Q_i,q_i;t_0)=\int {\cal D}Q{\cal D}q
\;\exp\big (\frac{i}{\hbar}{\cal S}[\dot Q,Q;\dot q,q]\big ) 
\;\;, \end{equation} 
with the boundary conditions $Q(t_i)=Q_i$, $q(t_i)=q_i$, $Q(t_f)=Q_f$, and $q(t_f)=q_f$, and 
where the superaction ${\cal S}$, corresponding to the super-Hamiltonian ${\cal H}'$ above, 
is defined as follows: 
\begin{eqnarray}\label{superaction}  
{\cal S}&\equiv&\int_{t_0}^t\mbox{d}\tau\;\big ({\cal T}(\dot Q,\dot q)-{\cal V}(Q,q)\big ) 
\\ [1ex] \label{superaction1} 
&:=&\int_{t_0}^t\mbox{d}\tau\;\Big ({\textstyle \frac{m}{2}}\dot Q^2-V(Q)
-\big ({\textstyle \frac{m}{2}}\dot q^2-V(q)\big )
-{\cal E}(Q,q)\Big )
\;\;. \end{eqnarray}
We recall that ${\cal E}\equiv 0$ corresponds to 
evolution according to the von\,Neumann equation, whereas ${\cal E}\neq 0$ represents 
classical dynamics in accordance with the Liouville equation, 
cf. Eqs.\,(\ref{Schroed})--(\ref{I}). 

However simple this result may seem, the Eqs.\,(\ref{superpropagator})--(\ref{superaction1}) 
present a new approach to describe time evolution of the full density matrix, with the 
particular feature that classical and quantum mechanical motion are formally treated  
in parallel, differing only in the action entering the phase in the 
path integral~\footnote{We remark that a complementary approach, based on the 
effective action generating equal-time correlation functions for 
nonequilibrium statistical systems, has been presented 
in Ref.\,\cite{Wetterich}, which results in evolution equations 
for a truncated set of correlation functions.}.  

We emphasize that our derivations are not confined to one-dimensional 
or single-particle physics, but can be extended as well all the way to relativistic 
field theories. Various applications come to mind here, some of which will be discussed in 
the following and in the concluding section.    

\subsection{Perturbation theory and superpropagator Dyson equation}
Considering the splitting of the superaction as in Eq.\,(\ref{superaction}), the 
perturbation theory naturally departs from organizing contributions to the full 
superpropagator, Eq.\,(\ref{superpropagator}), according to powers of the 
``perturbation'' ${\cal V}$. Sometimes it may be advantageous to include parts 
of the perturbation into the ``free'' part ${\cal T}$. This must be familiar from 
quantum mechanics, which presents a special case of our general considerations here. 
 
To begin with, if ${\cal V}(Q,q)\equiv V(Q)-V(q)$, corresponding to the superoperator 
related to the von\,Neumann equation, then the path integrals in Eq.\,(\ref{superpropagator}) 
factorize and we recover quantum mechanics. 
   
In the absence of an external potential or other interactions 
(${\cal V}\equiv 0$), the zeroth order or {\it free superpropagator} ${\cal G}_0$ is obtained 
as: 
\begin{eqnarray}\label{G0}
{\cal G}_0(Q_f,q_f;t|Q_i,q_i;t_0)&=&\int {\cal D}Q{\cal D}q
\;\exp\big (\frac{i}{\hbar}\int_{t_0}^t\mbox{d}\tau\;{\cal T}(\dot Q,\dot q)\big )
\\ [1ex] \label{G01} 
&=&G_0(Q_f,t;Q_i,t_0)G_0^\ast (q_f,t;q_i,t_0)
\;\;, \end{eqnarray} 
in terms of the well known free quantum mechanical propagator $G_0$, cf. 
Eqs.\,(\ref{piprop}})--({\ref{pathintegral2}), which is explicitly given by~\cite{Schulman}: 
\begin{equation}\label{G0QM} 
G_0(x,t;y,t_0)\equiv G_0(x,y;T:=t-t_0)
=\big (\frac{m}{2\pi i\hbar T}\big )^{1/2}\exp\Big (\frac{im}{2\hbar T}(x-y)^2\Big ) 
\;\;, \end{equation} 
for a free nonrelativistic particle of mass $m$. Remarkably, this 
zeroth order result is {\it identical} for classical and quantum mechanical 
propagation. 

Despite the fact that the free propagator for the Schr\"odinger equation 
incorporates such phenomena as the quantum mechanical spreading of a wave packet, 
we learn here that it also describes the propagation of a {\it free classical 
particle}. It is straightforward to verify -- following the transformations 
between Eqs.\,(\ref{HamiltonianF}) and Eq.(\ref{I}) -- that a massive particle, 
initialized as $\rho (x,p,t_0):=2\pi \delta (x-x_0)\delta (p-p_0)$ is propagated 
to $\rho (x,p,t):=2\pi \delta (x-x_0-Tp/m)\delta (p-p_0)$, as expected.   

For the perturbative expansion, we employ the standard formula: 
\begin{eqnarray}
&\;&\exp\big ( -\frac{i}{\hbar}\int_0^t\mbox{d}\tau\;{\cal V}(Q,q)_\tau \big )= 
\sum_{n=0}^\infty\frac{1}{n!}\big (\frac{-i}{\hbar} 
\int_0^t\mbox{d}\tau\;{\cal V}(Q,q)_\tau\big )^n 
\\ [1ex] 
&=&
\sum_{n=0}^\infty\big (\frac{-i}{\hbar}\big )^n  
\int_0^t\mbox{d}\tau_1\;{\cal V}(Q,q)_{\tau_1}\;\dots\;\int_0^{\tau_{n-1}}\mbox{d}\tau_n\;
{\cal V}(Q,q)_{\tau_n} 
\;\;, \end{eqnarray} 
with ${\cal V}(Q,q)_{\tau_k}:= {\cal V}\big (Q(\tau_k),q(\tau_k)\big )$.   

In order to analyze such terms at a given order, we make use of the important 
semigroup property of the (free) propagator and obtain to first order 
in the perturbation~\cite{FabThesis}: 
\begin{eqnarray}
&\;&{\cal G}(Q,q;t|Q',q';t_0)={\cal G}_0(Q,q;t|Q',q';t_0) 
\nonumber \\ [1ex] \label{firstorder} 
&\;&-\frac{i}{\hbar}\int_{t_0}^t\mbox{d}\tau\int\mbox{d}x\mbox{d}y\;{\cal G}_0(Q,q;t|x,y;\tau )
{\cal V}(x,y){\cal G}_0(x,y;\tau |Q',q';t_0)+\mbox{O}({\cal V}^2) 
\;, \end{eqnarray} 
to be illustrated explicitly by the result for an anharmonic potential shortly. 

We remark that on the right-hand side of Eq.\,(\ref{firstorder}) the superpotential 
is preceded (and followed) by a zeroth order propagator. This observation, which similarly 
holds at every order of this expansion, leads to a recursion relation of 
the $k$-th order propagator in terms of the $(k-1)$-th order one. 
This allows us to resum the perturbation series in the form of    
a {\it Dyson integral equation} for the full superpropagator: 
\begin{eqnarray}
&\;&{\cal G}(Q,q;t|Q',q';t_0)={\cal G}_0(Q,q;t|Q',q';t_0) 
\nonumber \\ [1ex] \label{Dyson} 
&\;&-\frac{i}{\hbar}\int_{t_0}^t\mbox{d}\tau\int\mbox{d}x\mbox{d}y\;{\cal G}_0(Q,q;t|x,y;\tau )
{\cal V}(x,y){\cal G}(x,y;\tau |Q',q';t_0) 
\;\;. \end{eqnarray} 
The whole procedure follows the usual one in quantum mechanics, yet includes the 
case of classical mechanics, and possibly others, for a suitably chosen superaction.   

\subsection{Illustration: the case of an anharmonic potential}

In order to make our general derivations more concrete and to extract some interesting 
general aspects, it may be useful to consider the example of a massive particle in 
an anharmonic potential, $V(x):=\lambda x^4$, where $\lambda$ is the coupling constant. -- 
We recall that for constant, linear, or harmonic coupling terms there is no difference 
between classical and quantum dynamics, cf. (\ref{Ezero}), in the representation 
that we have developed in this article. 

The calculations evaluating the superpropagator to first order, here with: 
\begin{equation}\label{anharmonic} 
{\cal V}(x,y)\equiv (x-y)V'\big (\frac{x+y}{2}\big ) 
=\frac{\lambda}{2}\big (x^4-y^4+2(x^3y-xy^3)\big ) 
\;\;, \end{equation} 
for classical dynamics (${\cal V}(x,y)\equiv V(x)-V(y)$ for quantum mechanics) consist in straightforward (if tedious) multiple Gaussian integrals, 
according to Eqs.\,(\ref{G0})--(\ref{G0QM}) and Eq.\,(\ref{firstorder}).  
The final result is: 
\begin{eqnarray}
&\;&{\cal G}(Q,q;t|Q',q';t_0)\;=\;{\cal G}_0(Q,q;t|Q',q';t_0) 
\nonumber \\ [1ex] \label{anharmonicprop1} 
&\;&\;\;\cdot\; 
\Big (1-\frac{i}{\hbar}\lambda\big [
C_1\Gamma_{\mbox{QM}}(Q,q;Q',q';T)
+C_2\Gamma_{\mbox{CL}}(Q,q;Q',q';T)\big ]\Big ) 
+\mbox{O}(\lambda^2)   
\;, \end{eqnarray} 
where $T:=t-t_0$, the coefficients for classical dynamics, 
$C_1:=1/2,\;C_2:=1/2$ ($C_1:=1,\;C_2:=0$ for quantum mechanics), 
and with the function:
\newcommand{\Qp}{{Q'}}
\newcommand{\qp}{{q'}}
\begin{eqnarray}
&\;&\Gamma_{\mbox{QM}}(Q,q;Q',q';T):=\frac{T}{5}\Big [
\frac{1}{2}\frac{i\hbar T}{m}(3Q^2+4Q\Qp +3\Qp^2)
\nonumber \\ [1ex] \label{GammaQM}
&\;&\;\;+Q^4+Q^3\Qp +Q^2\Qp^2+Q\Qp^3+\Qp^4 
\Big ] 
-\frac{T}{5}\Big [(Q,\Qp )\longleftrightarrow (q,\qp )\Big ]^\ast 
\;\;, \end{eqnarray} 
where the term is repeated, as indicated, with an exchange  
of variables and complex conjugation. Similarly: 
\begin{eqnarray}
&\;&\Gamma_{\mbox{CL}}(Q,q;Q',q';T):=\frac{T}{5}\Big (\;
\frac{i\hbar T}{m}\big (3Qq+2Q\qp +2\Qp q+3\Qp\qp\big )
\nonumber \\ [1ex] 
&\;&\;\;+\frac{1}{2}\big [Q^3(4q+\qp )+Q^2\Qp (3q+2\qp )+Q\Qp^2(2q+3\qp )+\Qp^3(q+4\qp )\big ] 
\nonumber \\ [1ex] \label{GammaCL}
&\;&\;\;-\frac{1}{2}\big [(Q,\Qp )\longleftrightarrow (q,\qp )\big ]\;\Big )
\;\;. \end{eqnarray} 

This result shows several interesting features. First of all, the perturbative expansion 
turns out to be a short-time expansion, with the overall scale of the first order 
correction set by $\lambda T$. Furthermore, different contributing terms differ by a scale 
set by $T/m$, i.e., by $T\;\times$\;Compton wavelength of the particle. Numerical 
studies visualizing the outcome here are presently underway~\cite{FabThesis}. 

However, most interesting seem general similarities and differences between classical (``CL'')
and quantum mechanical (``QM'') result in Eq.\,(\ref{anharmonicprop1}). The CL result 
has the same zeroth order term as QM; at first order, CL has one term  
in common with QM which, however, is reduced by an overall factor 1/2. This obviously stems 
from the varied expressions for ${\cal V}$ between CL and QM, cf. Eq.\,(\ref{anharmonic}). 
For the same reason, CL has additional terms, collected in  
$\Gamma_{\mbox{CL}}$, Eq.\,(\ref{GammaCL}), which are absent in QM.  

\subsection{Intra- and inter-space entanglement}

There is a {\it qualitative difference between CL and QM}, contained in $\Gamma_{\mbox{CL}}$  
and based on the different superoperators that enter the full path integral, 
Eqs.\,(\ref{superpropagator})--(\ref{superaction1}). Equivalently, since the QM evolution is 
generated by a commutator of the Hamiltonian with the density operator $\hat\rho$, it 
superposes and, 
for multi-partite systems, generally, entangles underlying 
bra- and ket-states separately, $\propto H_{ij}\rho_{jk}-\rho_{ij}H_{jk}$. For a bi-partite 
system, it is revealing to write such terms more clearly as: 
\begin{equation}\label{QMentangle}
[\hat H_{int},\hat\rho ]=\hat H_1\hat\rho_1\otimes\hat H_2\hat\rho_2-\hat\rho_1\hat H_1 
\otimes\hat\rho_2\hat H_2 
\;\;, \end{equation} 
for an interaction $\propto \hat H_1\otimes\hat H_2$, with the factors acting on 
subsystems ``1'' and ``2'', respectively, and where $\hat\rho =\hat\rho_1\otimes\hat\rho_2$,   
for a separable initial state. 
This has been called {\it dynamically assisted entanglement generation}, see, for example, Refs.\,~\cite{Jacquod,JacquodRev,Hornberger}. 

It may come 
as a surprise that the CL evolution does this just as well, 
due to the contribution of $\Gamma_{\mbox{QM}}$ for the first two terms on the right-hand 
side of Eq.\,(\ref{anharmonic}}) or, generally, due to the superoperator ${\cal E}$ 
of our earlier considerations. For polynomial interactions, for example, this superoperator 
{\it always} contains a contribution proportional to the usual QM terms. 

However, the CL evolution produces additional correlations in $\hat\rho$, due to the 
generator $\propto {\cal L}_{ij;kl}\rho_{kl}$, 
which possibly {\it entangles bra- and ket-states}. -- In comparison with Eq.\,(\ref{QMentangle}), 
for example, such terms can have the unfamiliar structure: 
\begin{equation}\label{CLentangle} 
\hat H'_1\hat\rho_1\otimes\hat\rho_2\hat H'_2-\hat\rho_1\hat H'_1\otimes\hat H'_2\rho_2 
\;\;, \end{equation} 
which differs decidedly from a commutator. -- This leads us to distinguish {\it intra-} 
(i.e., within given tensor product Hilbert space of subsystems ``1'' and ``2'') and 
{\it inter-space entanglement} (i.e., between said Hilbert space and its dual).  

For example, consider the anharmonic potential 
$V(x_1-x_2):=\lambda (x_1-x_2)^4$ for a bi-partite system consisting of particles ``1'' 
and ``2''. Following and suitably generalizing our derivation in Section\,2, 
this leads to the interaction: 
\begin{equation}\label{2particle} 
{\cal V}(Q_1,Q_2;q_1,q_2)=
\frac{1}{2}\lambda\big (Q_1-q_1-(Q_2-q_2)\big )\big (Q_1+q_1-(Q_2+q_2)\big )^3 
\;\;, \end{equation} 
in terms of variables introduced previously, taking into account both subsystems;  
similarly as before, the $Q$ and $q$ variables refer to bra- and 
ket-states, respectively. Besides the separable terms, 
$\propto (Q_a-q_a)(Q_a+q_a)^3,\; a=1,2$, there are the terms which mix (and entangle) 
variables of both subsystems, as usual in QM. However, there are clearly additional 
terms  
that refer to Hilbert space and its dual simultaneously (and entangle corresponding 
states), for example, $\propto Q_aQ_bq_b^2,\; b\neq a$.   

In retrospect, somehow, such difference between CL and QM 
evolution had to be expected:  
instead with superstates $|Q,q\rangle\rangle$, we could have 
worked with superstates $|x,p\rangle\rangle$, relating to coordinates and momenta of 
the classical theory. There, coordinates and momenta end up tightly correlated, due to 
Hamilton's equations, and produce inter-space entanglement in an interacting bi-partite 
system. 

Thus, we find that the confrontation of CL with QM, as in our side-by-side study, 
is quite revealing. In particular, we speculate that this opens new views on  
generating entanglement in multipartite systems, perhaps, by evolving through 
quasiclassical stages or by making use of decohered intermediary 
states~\footnote{Previous considerations of the semiclassical regime, such as in 
Refs.~\cite{Jacquod,JacquodRev}, were motivated as suitable approximations of the 
quantum mechanical evolution, in particular, for studies of the different decoherence 
properties between classically regular and chaotic systems. Our results seem 
to show that crossing the quantum-classical divide may offer an additional resource 
for entanglement generation and related ``truly quantum'' phenomena. This might be related to   
the ``underlying reality'' of (CL and QM) physics, assumed in Refs.\,\cite{Zwitters,Khrennikov}, 
consisting in statistical correlations.}.   

Concerning the quantum-classical divide, the present 
analysis shows that there is a deep formal similarity between CL and QM. However, 
this also demonstrates that what has been discussed in various ways as CL limit 
of QM -- and which is similarly relevant for ``emergent QM'' -- deserves more 
study. 

While our work has been concerned mainly with the evolution of CL or QM objects, 
we recall that V.I.\,Man'ko and collaborators have pointed out that classical 
states may differ widely from what could be obtained as the ``$\hbar\rightarrow 0$'' 
limit of quantum mechanical ones. They show that all states can 
be classified by their `tomograms' as {\it either} CL {\it or} QM, CL {\it and} QM, 
and {\it neither} CL {\it nor} CM~\cite{Manko}.  

The classical limit might be a ``{\it F}or{\it A}ll{\it P}ractical{\it P}urposes'' 
limit, gradually approached through decoherence or ``$\hbar\rightarrow 0$''. However, 
in order to bridge (if at all) the qualitative difference between intra- and inter-space 
entanglement that we find, and explain the ``Man'ko classes of states'', 
some unknown dynamics beneath still awaits to be 
uncovered~\footnote{A simple attractor model, motivated by assumptions about effects 
of fundamental spacetime discreteness~\cite{Elze09}, has been discussed in 
Ref.~\cite{ElzeAttractor}.}.   

\section{The {\it almost classical} Jaynes-Cummings model}

In this section, we apply our operator approach for the Liouville equation 
to a field theory, namely to a Rydberg atom interacting with the electromagnetic 
field. Following the approximations that lead to the quantum mechanical Jaynes-Cummings  
model~\cite{JaynesCummings}, we will show that the dynamics of this celebrated model is 
almost of classical character. As we shall see, if it were not for the anharmonic 
Coulomb interaction between electron and atom, the dynamics would be entirely classical.  

\subsection{The classical model}     
  
We consider an electron (mass $m$) interacting electromagnetically with a positive charge 
(atom) fixed at the origin and with the radiation field. Thus, we depart from the 
classical Lagrangian: 
\begin{equation}\label{L} 
L:=\frac{m}{2}\dot x^2+\int\mbox{d}^3r\;\big\{\frac{1}{8\pi}(E^2-B^2)-\rho\phi +J\cdot A\big\} 
\;\;, \end{equation} 
where the electric and magnetic fields, respectively, are given by: 
\begin{equation}\label{fields} 
E=-\dot A-\nabla\phi \;\;,\;\;\; B=\nabla\times A 
\;\;, \end{equation} 
as usual, in terms of vector and scalar potential, $A$ and $\phi$, respectively. The 
charge and current densities, $\rho$ and $J$, respectively, are given by: 
\begin{equation}\label{chargecurrent} 
\rho (r)=-e\delta^3 (r-x)+\delta^3 (r)\;\;,\;\;\; J(r)=-e\dot x\delta^3 (r-x) 
\;\;. \end{equation} 

Next, we introduce Fourier modes of the fields, with $A(k)=A^*(-k)$ and $\phi (k)=\phi^*(-k)$, 
since the fields are real. We choose the Coulomb gauge by imposing $A_\parallel (k)=0$, 
which implies $\nabla\cdot A=0$. Correspondingly rewriting the Lagrangian, we determine 
the canonical momenta, in order to obtain the Hamiltonian of the classical model: 
\begin{eqnarray}
H&=&\frac{1}{2m}\Big (p+e\int\mbox{d}^3k\; 
\big\{ A(k)\mbox{e}^{ik\cdot x}+A^*(k)\mbox{e}^{-ik\cdot x}\big\}\Big )^2-\frac{e^2}{|x|} 
\nonumber \\ [1ex] \label{JCHamiltonian} 
&\;&+\frac{1}{8\pi}\int\mbox{d}^3k\;\big\{\Pi^*(k)\cdot\Pi (k)+k^2A^*(k)\cdot A(k)\big\} 
\\ [1ex] \label{JCHamiltonian1} 
&\equiv&H(x,p;A,\Pi^*;A^*,\Pi ) 
\;\;, \end{eqnarray} 
where we indicate the canonically conjugated pairs of variables of the Hamiltonian; 
the momentum integrations have to take into account that not all Fourier modes are 
independent, for real fields.  

We are now in the position, cf. Section\,2, to describe this model in phase space. 
We proceed in four steps: 
\begin{itemize}
\item First, we introduce the probability density (over phase space) 
$\rho (x,p;A,\Pi^*;A^*,\Pi )$, which will be interpreted, as before, as matrix element 
of a Hermitean density operator $\hat\rho$. We assume that the atom-electromagnetic-field 
system is confined to a cavity of finite volume $V$, thus replacing integrals by discrete 
mode sums, $\int\mbox{d}^3k\;g(k)\rightarrow V^{-1}\sum_k\;g_k$.  
\item Second, we obtain the Liouville equation, $-\partial_t\rho =\{ H,\rho\} =\dots\;$, 
evaluating the relevant Poisson bracket. 
\item Third, we replace momenta by coordinates via Fourier transformation, $p\rightarrow y$, 
$\Pi_k^*\rightarrow B_k$, $\Pi_k\rightarrow B_k^*$. 
\item Fourth, we perform the "Wigner rotations", $Q:=x+y/2$, $q:=x-y/2$, $Q_k:=A_k+B_k/2$, 
and $q_k:=A_k-B_k/2$. 
\end{itemize}
Details and the following derivations will be reported elsewhere~\cite{GioThesis}. 
  
Introducing the following notation: 
\begin{equation}\label{potentials}
V(\chi ):=-\frac{e^2}{|\chi |}\;\;,\;\mbox{for}\;\;\chi =Q,q\;\;; 
\;\;\;{\cal E}(Q,q):=4e^2\frac{Q^2-q^2}{|Q+q|^3}-V(Q)+V(q)
\;\;, \end{equation}  
where we suppress constant normalization factors etc., as before, we 
obtain the remarkable result that the  
{\it classical} evolution equation is: 
\begin{eqnarray}\label{rhoevol} 
i\partial_t\rho &=&\big\{\mbox{``von Neumann''}+{\cal E}+\Gamma +\Sigma \big\}\rho 
\\ [1ex]  
&\equiv&
\Big\{ -\frac{1}{2m}\partial_Q^{\; 2}+V(Q)-\big (-\frac{1}{2m}\partial_q^{\; 2}+V(q)\big ) 
\nonumber \\ [1ex] 
&\;&\;+\frac{1}{8\pi}\sum_k\big [
-\partial_{Q_k}\cdot\partial_{Q_k^*}+\omega_k^{\; 2}Q_k\cdot Q_k^*
-\big (-\partial_{q_k}\cdot\partial_{q_k^*}+\omega_k^{\; 2}q_k\cdot q_k^*\big )\big ] 
\nonumber \\ [1ex] 
&\;&\;-i\frac{e}{m}\sum_k\big [ 
\mbox{e}^{ik\cdot Q}Q_k\cdot\partial_Q+\mbox{e}^{ik\cdot q}q_k\cdot\partial_q\big ] 
\nonumber \\ [1ex] 
&\;&\;+\frac{e^2}{2m}\sum_{k,k'}\big [ 
Q_k\cdot Q_{k'}^*\mbox{e}^{i(k-k')\cdot Q}-q_k\cdot q_{k'}^*\mbox{e}^{i(k-k')\cdot q}\big ]
\Big\}\rho 
\nonumber \\ [1ex] \label{rhoevol1} 
&\;&+\Big\{ {\cal E}(Q,q)+\Gamma +\Sigma \Big\}\rho 
\;\;, \end{eqnarray}
with $\rho\equiv\rho (Q,Q_k,Q_k^*;q,q_k,q_k^*;t)$, where $k$ runs over all modes,  
$\omega_k:=|k|$, and where $\Gamma$ and $\Sigma$ denote rather complicated 
terms that involve all phase space variables; they are given explicitly in 
Ref.\,~\cite{GioThesis}. While the last line of Eq.\,(\ref{rhoevol1}) presents additional 
terms, in particular the superoperator ${\cal E}$, the previous terms represent exactly the 
terms of the {\it quantum mechanical} von\,Neumann equation for the atom-field system under 
consideration; besides further interaction terms, due to minimal coupling, we find the 
contribution of the free electromagnetic field in the second and those of the electron 
interacting with the Coulomb potential of the Rydberg atom in the first line, respectively. 

We anticipate that in the {\it dipole approximation} we have $\Gamma,\Sigma\rightarrow 0$. 
Therefore, we do not study further the impact of those terms here~\cite{GioThesis}. 

Instead, we recall the well known additional approximations that turn the von\,Neumann terms 
of Eqs.\,(\ref{rhoevol})--(\ref{rhoevol1}) into those of the 
{\it Jaynes-Cummings model}~\cite{JaynesCummings}:
\begin{itemize}
\item The {\it dipole approximation}, assuming that $\tilde k\cdot l\ll 1$, where 
$\hbar\tilde k$ and $l$ denote a typical photon momentum and linear size of a Rydberg 
electron orbit, respectively.  
\item The restriction to {\it one cavity photon mode} with energy $\hbar\omega$. This 
yields the approximate Hamilton operator~\cite{GioThesis}: 
$$\hat H=\sum_i\; \omega_i|i\rangle\langle i|+\omega (\hat a^\dagger \hat a+\frac{1}{2}) 
+i\sum_{i\neq j}d_{ij}(\hat a-\hat a^\dagger )|i\rangle\langle j| \;\;, $$ 
where the sums run over the Rydberg levels, with energies $\hbar\omega_i$, 
$\hat a^{(\dagger )}$ are photon annihilation 
(creation) operators, and where the last term involves the dipole transition amplitudes 
$d_{ij}$.  
\item The restriction to a {\it two-level subspace}, spanned by states $|g\rangle,|e\rangle$; 
the lower level energy is conveniently set to $\hbar\omega_g\equiv 0$, while the physical 
realizations considered, usually, have $\omega_e\approx\omega$, i.e., approximately resonant 
photon and excited electron states.  
\item The {\it rotating wave approximation}, which yields the `energy conserving' 
dipole interaction term 
$\hat{\cal D}\propto\hat a|e\rangle\langle g|-|g\rangle\langle e|\hat a^\dagger$. 
\end{itemize}
The resulting Jaynes-Cummings Hamiltonian is: 
\begin{equation}\label{JCH} 
\hat H_{\mbox{JC}}=\omega_e|e\rangle\langle e|+\omega (\hat a^\dagger \hat a+\frac{1}{2}) 
+id_{eg}\big (\hat a|e\rangle\langle g|-|g\rangle\langle e|\hat a^\dagger\big )
\;\;. \end{equation}  
Then, following the above derivation, the evolution equation becomes: 
\begin{equation}\label{JCevol} 
i\partial_t\hat\rho = [\hat H_{\mbox{JC}},\hat\rho ]+\hat{\cal E}\rho  
\;\;, \end{equation} 
where we appropriately incorporated here  
the superoperator $\hat{\cal E}$. This term presents the {\it only} difference between the  
{\it classical} dynamics described by Eqs.\,(\ref{rhoevol})--(\ref{rhoevol1}) and the usual  
{\it quantum mechanical} one. 

Thus, we find in this `standard model' of quantum optics a detailed example for 
the similarity between CL and QM evolution laws. 

\subsection{Dipole interaction and Coulomb superoperator as perturbations}

In order to illustrate our findings, we will briefly study the influence 
of the classical superoperator on the evolution described by the Jaynes-Cummings model 
in perturbation theory, while a more complete analysis will be presented 
in Ref.~\cite{GioThesis}. 

Following Section\,4.3, cf. Eq.\,(\ref{Dyson}), we presently treat the dipole interaction, 
$\hat{\cal D}\propto\hat a|e\rangle\langle g|-|g\rangle\langle e|\hat a^\dagger$,  
together with the superoperator $\hat{\cal E}$ of Eq.\,(\ref{JCevol}) as perturbation.  
Correspondingly, we choose the Rydberg atom states and one-mode photon number states for  
the tensor product basis of the relevant Hilbert space. 

Then, the density matrix evolves according to: 
\begin{equation}\label{JCEvol} 
\hat\rho (t)=\hat{\cal G}(t)\hat\rho (0) 
=\hat{\cal G}_0(t)\hat\rho (0)-i\int_0^t\mbox{d}\tau\;  
\hat{\cal G}_0(t-\tau )\big (\hat{\cal D}+\hat{\cal E}\big )\hat{\cal G}_0(\tau )\hat\rho (0) 
\;\;, \end{equation} 
to first order in $\hat{\cal D}+\hat{\cal E}$. While the dipole operator 
acts on atom and electomagnetic field states simultaneously, the superoperator acts only on 
the Rydberg states. Hence, the matrix elements of $\hat{\cal E}$ are defined by: 
\begin{equation}\label{Ematrix} 
{\cal E}_{ab,cd}:=\int\mbox{d}^3Q\mbox{d}^3q\;\psi^*_a(Q)\psi_b(q){\cal E}(Q,q)
\psi_c(Q)\psi^*_d(q) 
\;\;, \end{equation} 
where $\psi_{i\equiv (n,l,m)}$ denote standard hydrogen-like wave functions. 
Since these are eigenstates of parity, $P_i=\pm 1$, we find 
${\cal E}_{ab,cd}=P_aP_bP_cP_d{\cal E}_{ab,cd}$, which implies the selection rule:  
${\cal E}_{ab,cd}=0$, if $P_aP_bP_cP_d=-1$. Furthermore, we have:  
${\cal E}_{ab,cd}=-{\cal E}_{ba,dc}^*$. 

Consequently, the only nonzero matrix elements are 
${\cal E}_{eg,eg}$, ${\cal E}_{ge,ge}$, ${\cal E}_{ee,gg}$, and ${\cal E}_{gg,ee}$.  
The latter two vanish for the specific ground and excited states used in cavity QED 
experiments~\cite{CavityQED}.    

Taking matrix elements of Eq.\,(\ref{JCEvol}), we find that the superoperator 
only affects the evolution of the matrix elements $\rho_{eg|nn'}$ and 
$\rho_{ge|nn'}=\rho_{eg|n'n}^*$ (by hermiticity);    
here the Fock states are labelled by photon numbers $n,n'$. Then, we find: 
\begin{eqnarray} 
\rho_ {eg|nn'}(t)&=&\mbox{e}^{-it[\omega_e-\omega_g+\omega (n-n')]}
\Big ([1+it{\cal E}_{eg,eg}]\rho_{eg|nn'} (0)
\nonumber \\ [1ex] \label{rho1st} 
&\;&+d_{eg} t\big [\sqrt{n+1}\rho_{gg}(0)\rho_{n+1\;n'}(0)
-\sqrt{n'}\rho_{ee}(0)\rho_{n\;n'-1}(0)\big ]\Big )
\;\;, \end{eqnarray}
where we assume that the initial state factorizes, for simplicity. 

Thus, to first order in this perturbative expansion, we find that the superoperator 
competes with the dipole interaction, as far as the atom states are concerned; 
however, it does so without affecting the field states. Numerical estimates 
indicate that its matrix elements are not small compared to the ones of the dipole 
operator for cavity QED experiments (see Ref.~\cite{GioThesis} for further 
details)~\cite{CavityQED}. 

To summarize, 
{\it almost} all dynamical (operator) features of the Jaynes-Cummings model 
can be derived in the classical framework, as we have shown. 

Nevertheless, there {\it is} a noticeable difference between 
classical and quantum evolution in the version of the Jaynes-Cummings model that  
is related to cavity QED experiments. This is solely due to the classical superoperator, 
which stems from the {\it Coulomb interaction} between electron and Rydberg ion. 

In distinction, 
had we considered a charged particle trapped by a {\it linear or harmonic potential}, 
then the superoperator would vanish identically, 
cf. Section\,2, Eq.\,(\ref{Ezero}). The correspondingly 
modified Jaynes-Cummings model could be seen as of entirely classical 
origin, despite its quantum mechanical appearance. 

\section{Conclusions}

Beginning with the Liouville equation of classical statistical mechanics, 
we have introduced a (super)operator formulation~\cite{superspace}, 
which brings it as close as possible to 
the von\,Neumann equation of quantum mechanics, provided suitable coordinates 
are chosen in superspace~\cite{Elze05,ElzeAttractor}. Presently, we have concentrated 
on the similarities and differences between both evolution equations. 

We have chosen the Jaynes-Cummings model~\cite{JaynesCummings,CavityQED}, 
in particular, to illustrate both 
aspects and to show that this benchmark model of 
quantum optics and cavity QED can be interpreted to a large extent in terms 
of classical dynamics~\cite{GioThesis}. Furthermore, this model serves as an 
example that our more general considerations apply not only to single- or 
few-particle systems, but to field theories as well. 

While presently the relevant Hilbert space has been treated as tensor 
product space of the Rydberg electron single-particle and the photon Fock space, 
earlier also a functional approach combining fermion and boson 
fields has been discussed~\cite{Elze07}. 

More generally, we discussed in parallel the formal solutions of the Liouville and 
the von\,Neumann equations. Introducing suitable propagators, we derived a path 
integral representation for both cases side by side~\cite{FabThesis}. 

The path integral for the propagator of the Liouville equation is new and may have 
interesting applications in classical physics. We derived and illustrated 
the related perturbation theory. 

We discussed how these results and the action entering this path integral, 
in particular, hint at the possibility that a form of entanglement is also 
generated by classical dynamics, which has gone unnoticed before. 
It combines the quantum mechanical 
dynamically assisted entanglement generation~\cite{Jacquod,JacquodRev,Hornberger} 
with a classical counterpart. 
We call the former {\it intra-space entanglement}, since it acts separately within 
the Hilbert spaces of bra- and ket-states. In distinction, the classical 
dynamics additionally produces {\it inter-space entanglement}, i.e., it 
correlates the Hilbert space and its adjoint in addition to what would,  
otherwise, be recognized as quantum entanglement. 

If the relative strength of intra- and inter-space entanglement can be manipulated, 
for example, by driving a system dynamically between quantum and classical behaviour, 
this may open additional ways to influence the quantum mechanical entanglement, 
which is of central importance in research concerning quantum information and quantum 
foundations alike.    

The close relation between classical and 
quantum mechanical dynamics that we uncovered may help to address in new ways  
problems related to the nature of classical or quantum 
states~\cite{Zwitters,Manko,ElzeAttractor}, to the pathways, 
if any, over the quantum-classical divide~\cite{KieferEtAl,Zurek,tHooft06,Elze05,Vitiello01}, 
or the measurement problem~\cite{Schlosshauer,Adler}.     
  
\section*{Acknowledgments}

It is a pleasure to thank Nick Manton for several discussions of Liouville dynamics   
vs. QM, Vladimir Man'ko and Andreij Khrennikov for discussions of 
tomography of states and probabilistic formulations, and Marco Genovese for inviting 
H-T\,E to present this work at the 5th Workshop -- ad memoriam of 
Carlo Novero -- {\it ``Advances in Foundations of Quantum Mechanics and Quantum 
Information with Atoms and Photons''} (Torino, May 2010).  


\end{document}